# Particle Acceleration by Magnetic Reconnection in AGNs and in the IGM

**Elisabete M. de Gouveia Dal Pino[1]**
*Instituto de Astronomia, Geofísica e Ciências Atmosféricas (IAG-USP), Universidade de São Paulo*
*R. do Matão 1226, São Paulo, SP, Brazil*
*E-mail:* `dalpino@astro.iag.usp.br`

**Grzegorz Kowal**
*Instituto de Astronomia, Geofísica e Ciências Atmosféricas (IAG-USP), Universidade de São Paulo*
*R. do Matão 1226, São Paulo, SP, Brazil*
*E-mail:* `kowal@astro.iag.usp.br`

**Alex Lazarian**
*Astronomy Department, University of Wisconsin - Madison*
*Madison, WI, USA*
*E-mail:* `lazarian@wisc.edu`

**Reinaldo Santos-Lima**
*Instituto de Astronomia, Geofísica e Ciências Atmosféricas (IAG-USP), Universidade de São Paulo*
*R. do Matão 1226, São Paulo, SP, Brazil*
*E-mail:* `rlima@astro.iag.usp.br`

There is no single mechanism by which fast particles are accelerated in astrophysical environments, and it is now recognized that the data require a rich variety of different mechanisms operating under different conditions. The mechanisms discussed in the literature include varying magnetic fields in compact sources, stochastic processes in turbulent environments, and acceleration behind shocks. An alternative, much less explored mechanism so far, involves particle acceleration *within* magnetic reconnection sites. In this work, we explore this mechanism in the AGN framework and show that particles are efficiently accelerated through a first-order Fermi process and have an exponential growth of energy. We also address briefly the propagation of cosmic rays (CRs) in the intergalactic medium (IGM). Since the latter is a collisionless environment, kinetic effects must be considered which will affect the turbulent magnetic field distribution and therefore, the CR propagation.

---

[1] Speaker

# 1. Introduction

Acceleration of energetic particles is important for a wide range of astrophysical environments, from stellar magnetospheres, accretion disk/jet systems, supernova remnants and gamma ray bursts to clusters of galaxies. Several mechanisms for particle acceleration have been discussed in the literature which include varying magnetic fields in compact sources, stochastic second order Fermi process in turbulent interstellar and intracluster media, and the first order Fermi process behind shocks (see [1] for a review). An alternative, less explored mechanism so far, involves particle acceleration within magnetic reconnection sites. Here, we will discuss this mechanism in the framework of accretion disk/jet systems, with particular emphasis to the AGNs.

Magnetic reconnection occurs when two magnetic fluxes of opposite polarity encounter each other. In the presence of finite magnetic resistivity, the converging magnetic lines annihilate at the discontinuity surface and a current sheet forms there, releasing magnetic energy. Reconnection can be observed directly in the solar corona and flares, but can also be associated to a wide variety of astrophysical environments such as accretion disk/jet systems.

In 2005, de Gouveia Dal Pino & Lazarian [2] proposed a mechanism to accelerate particles to relativistic velocities *within* the reconnection zone in a similar way to the first-order Fermi process that occurs in shocks, i.e., they proposed that charged particles may bounce back and forth several times and gain energy due to head-on collisions with the two converging magnetic fluxes of opposite polarity that move to each other with the reconnection velocity ($v_{rec}$) carrying flow with them. They found that the particle enegy gain after each round trip is $\Delta E/E \propto v_{rec}/c$. Now, under *fast* reconnection conditions (see below), $v_{rec}$ is of the order of the local Alfvén speed ($v_A$) [3] and thus in the surruondings of relativistic sources $v_{rec} \sim v_A \sim c$. They have also shown that the accelerated particles have a power-law distribution $N(E) \propto E^{-5/2}$ and a corresponding electron synchrotron radio power-law spectrum $S_\nu \propto \nu^{-0.75}$ which is compatible, for instance, with the observed radio flares of galactic microquasars ([2] and references therein). Later, Drake et al. [4] appealed to a similar process, but within a collisionless reconnection scenario.

The acceleration process above was originally discussed in the context of microquasars and has been recently extended to astrophysical jet-accretion disks in general [5,6]. In these works, it has been argued that violent magnetic reconnection between the magnetic field lines arising from the inner disk region and those anchored in the central object are able to heat the coronal/disk gas and accelerate the plasma to relativistic velocities. In particular, in the case of relativistic systems, a diagram of the magnetic energy rate released by violent reconnection as a function of the black hole (BH) mass which spans $10^9$ orders of magnitude was derived and demonstrates that the magnetic reconnection power is more than sufficient to explain the observed radio outbursts, from microquasars to low luminous active galactic nuclei (AGNs).

In this work, we will review recent results of 2D and 3D MHD simulations that were carried out in order to test the acceleration model above occurring within reconnection sites.

## 2. Particle acceleration within magnetic reconnection sites: MHD simulations

In order to test the predictions of the acceleration model described in the previous section, we performed numerical simulations solving the isothermal magnetohydrodynamic (MHD) equations in two- and three-dimensions (2D and 3D, respectively) [7,8].

In the 2D approach, we built a set up of eight Harris current sheets in a periodic box with initial total (gas plus magnetic) pressure uniform (see also [9]). We imposed random weak velocity fluctuations to this environment in order to enable spontaneous reconnection events and the development of magnetic islands. We then injected test particles in a given snapshot of this MHD domain and integrated their trajectories solving the equation of motion for each charged particle $d(\gamma m\mathbf{u})/dt = q\,(\mathbf{u}-\mathbf{v}) \times \mathbf{B}$, where m, q and **u** are the particle mass, electric charge and velocity, respectively, **B** and **v** are the magnetic field and plasma velocity, respectively, and $\gamma$ is the Lorentz factor. We have considered only protons and, for simplicity, assumed the speed of light c to be 20 times the Alfvén speed. **Figure 1** presents an example of such evolved 2D configuration of the magnetic field structure with magnetic islands. We found that within the contracting magnetic islands and current sheets the particles accelerate exponentially predominantly through the first order Fermi process, as previously described, while outside the current sheets and islands, the particles experience mostly drift acceleration due to magnetic fields gradients. In Figure 1 configuration, if we do not consider an-out-of-plane guide field, the parallel component of the acceleration saturates at some level. When a guide field is included, this constraint is removed since particles can continue increasing their parallel speed as they travel along the guide field. In fully 3D MHD models, the acceleration of the particles exhibits the same trend as in 2D models with a guide field, i.e., there is no constraint on the acceleration of the parallel speed. The initial thermal energy distribution of the injected particles in the domain of Figure 1 quickly develops a kinetic energy spectrum with a tail in the high energy range. After $\sim 10^3$ hrs, the particle energy has increased up to $10^8$ orders of magnitude (see more details in [7]).

In more realistic 3D domains, we have explored three classes of models. In one case, we considered a single current sheet formed by the encounter between two large scale magentic fluxes of opposite polarity, as in the so called Sweet-Parker configuration; in a second model, we injected turbulence within the current sheet; and in a third model we considered a pure turbulent envirionment .

In the standard Sweet-Parker (S-P) model [10], the reconnection speed is given by $v_{rec} \approx v_A\, S^{-1/2}$, where $S = LV_A/\eta$, L is the length of the reconnection layer (see Figure 1), and $\eta$ is the Ohmic diffusivity. Because of the typical huge astrophysical sizes (L), S is also huge for Ohmic resistivity values ( e.g., for the ISM, $S \sim 10^{16}$). Thus, unless we invoke an anomalously large (non-Ohmic) resistivity, the S-P reconnection is very *slow*. However, observations indicate that magnetic reconnection must be *fast* in some circustances (e.g., solar flares). Vishniac & Lazarian [3] proposed a model for fast reconnection that independs on the resistivity. The model appeals to the ubiquitous astrophysical turbulence as a universal trigger of fast reconnection. The predictions of this model have been successfully tested in numerical simulations [11] which confirmed that the reconnection speed is of the order of the Alfvén speed. An important consequence of the fast reconnection of turbulent magnetic fields is the formation of a thick volume filled with reconnected small magnetic fluctuations. In order to test the acceleration of particles within 3D domains of fast reconnection induced by turbulence, we introduced turbulent flow within a current sheet formed by two large scale converging magnetic flux tubes (as in the S-P configuration; see **Figure 2**), then we followed the trajectories of test particles injected in this domain. We found that the presence of turbulence siginificantly increases the acceleration rate in a 1st-order Fermi process. The particles trapped within the current sheet suffer several head-on scatterings by the contracting magnetic fluctuations in the thick volume embedded in the current sheet, as shown in **Figure 3**. We have also performed 3D simulations of the acceleration of particles injected in a single S-P current sheet where the reconnection speed was made *artificially* fast by assuming a large resistivity η (which makes S small). In this case, we found that the acceleration rate is slightly smaller due to the thinner current sheet, but the process is still a 1st-order Fermi. For comparison, Figure 3 also shows the acceleration of particles in a pure turbulent 3D environment like the one we may find, for instance, in the ISM. In this case, as the particles suffer collisions both with approaching and receding magnetic irregularities the acceleration rate is much smaller and is a 2$^{nd}$ order Fermi process.

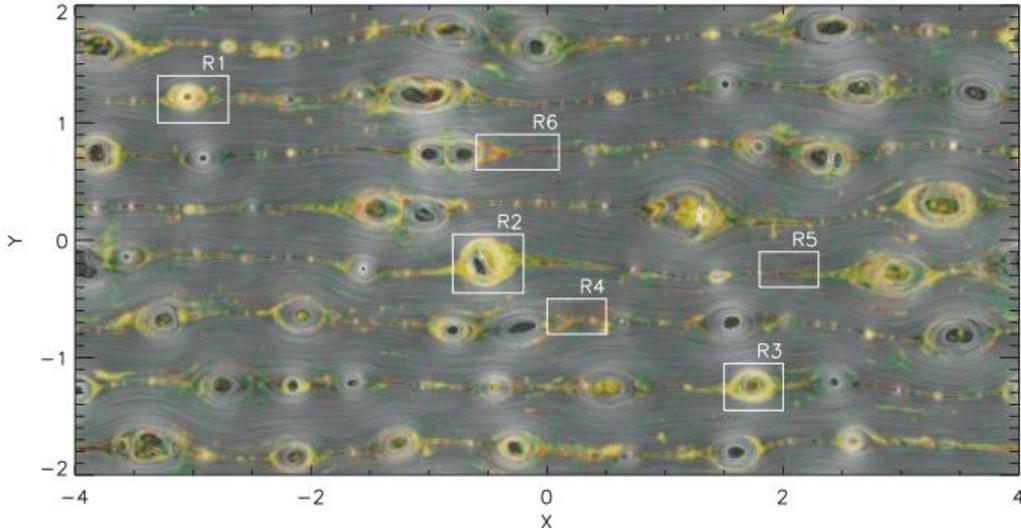

**Figure 1**. Topology of the magnetic field represented with a gray texture with semi-transparent color maps representing locations where the parallel and perpendicular particle velocity components are accelerated for a 2D model (with $B_z = 0.0$ at time 6.0 in the code units). The red and green colors correspond to regions where either parallel or perpendicular acceleration occurs, respectively, while the yellow color shows locations where both types of acceleration occur. The parallel component increases in the contracting islands and in the current sheets as well, while the perpendicular component increases mostly in the regions between current sheets. The simulation was performed with the resolution 8192x4096. We injected 10,000 test particles in this snapshot with the initial thermal distribution with a temperature corresponding to the sound speed of the MHD domain (extracted from [7]).

    **To summarize, the results just described** show that the acceleration within current sheets with turbulence can be extremely efficient and could be a powerful mechanism around BH/accretion disks at the jet launching region of AGNs, as represented in Figure 2 [8]. In forthcoming work we will also include the relevant loss mechanisms of the CRs in these soruces

in order to assess the importance of this acceleration mechanism in comparison to other processes (e.g., diffusive shock acceleration) in order to build realistic source light curves.

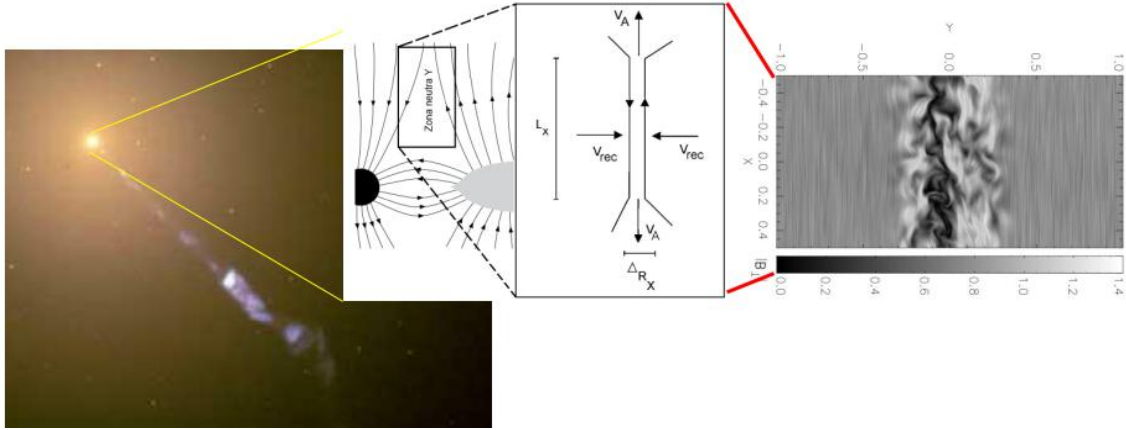

**Figure 2**. From left to right the figure shows: the HST image of M87 AGN; a schematic representation of the expected magnetic field structure around the accretion disk and the central BH (as in [2]); a schematic representation of the reconnection zone with the two converging magentic fluxes of opposite polarity as in a Sweet-Parker configuration [2]; and a 3D MHD simulation of magnetic reconnection with turbulence injected within the current sheet to make reconnection fast (as in [11,8]).

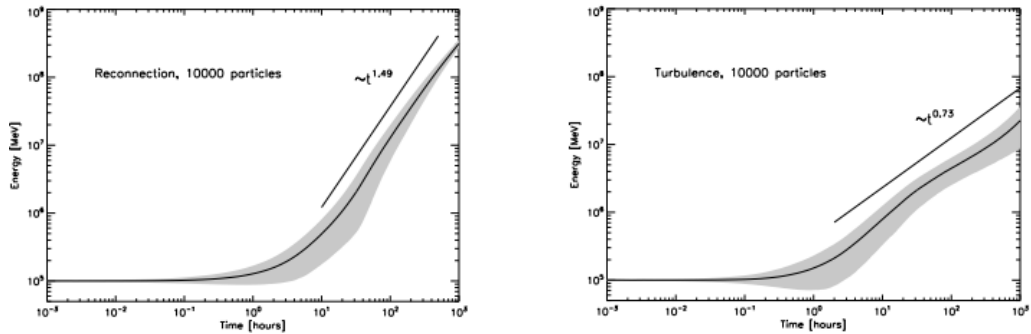

**Figure 3**. Comparison of particle acceleration rate (without losses included) within a current sheet with turbulence which makes the magenetic reconnection fast and enhances the efficiency of the 1$^{st}$ order Fermi process that takes place there (left); and in a pure MHD turbulent medium where a 2$^{nd}$ order Fermi process takes place (right). The left process could be occurring around the AGN accretion disks at the jet launching basis, while the right process is expected to be occuring in turbulent diffuse media like the ISM [8].

## 3. Cosmic ray propagation in the intergalactic and intracluster media

The propagation of cosmic rays (CRs) is conditioned by the interaction of these energetic particles with the turbulent magnetic fields that permeate the ISM and the intergalactic (IGM) or intracluster (ICM) media. However, the hot plasma in the ICM and IGM has very low density ($n \sim 10^{-3}$ cm$^{-3}$) which makes it weakly collisional, i.e., the Larmor frequency is much larger than the ion-ion collision frequency. For instance, in the Hydra A cluster, the ion Larmour radius is typically $\sim 10^5$ km, while the mean free path between collisions is $\sim 10^{15}$ km . In this case, a standard MHD description of the turbulent environment is not appropriate because the microscopic velocity distribution of the particles is not isotropic and gives rise to kinetic effects. The thermal pressure becomes anisotropic with respect to the magnetic field orientation and the evolution of the turbulent gas is more correctly described by a kinetic MHD (KMHD) approach. In the KMHD, the equations that describe the system evolution are the MHD equations, except for the thermal pressure term which is replaced by two components, one parallel and another perpendicular to the magnetic field. In recent work ([12,13] and references therein), 3D numerical KMHD simulations of the turbulent IGM were performed employing either isothermal or quasi-adiabatic formulations for the pressure tensor. These studies have shown that the pressure anisotropy may give rise to small-scale (mirror and firehose) instabilities which cause the kinetic and magnetic energies to accumulate in the smallest scales. This leaves the turbulent structure of the IGM much more *wrinkled* than in a standard MHD turbulent system. This may change substantially the current CR propagation paradigm in the IGM, particularly in the ultra-high energy range and may provide important new constraints to be assessed in near future by the CTA.